\documentstyle[12pt]{article}

\newcommand {\parn}{\par \noindent}
\newcommand{\num} {\hspace {-7 mm}\leqno}
\newtheorem{th}{\bf {Theorem }}
\newtheorem{lem}[th]{\bf {Lemma}}
\newtheorem{prop}[th]{\bf {Proposition}}

\newcommand\n{I\hskip-4pt N}

  \newcommand\r{I\hskip-3pt R}

\newcommand{\fta}{\Phi_{ \lambda , \omega}}

\newcommand{\Fta}{f_{ \lambda , \omega}}
\newcommand{\Pta}{(p f)_{ \lambda , \omega}}

\newcommand{\ftb}{\Psi_{\lambda , \omega}}

\newcommand{\Ftb}{g_{\lambda , \omega}}
\newcommand{\Ptb}{(p g)_{ \lambda , \omega}}

\newcommand{\ex}{e^{ i \sqrt{\lambda} x. \omega}}

\newcommand{\la}{{\sqrt{\lambda}}}
\newcommand{\T} { { {\tau} \over {\sqrt{\lambda}}}}

\begin{document}

\centerline{\large {\bf{An inverse scattering problem for short-range }}}
\centerline{\large {\bf{systems in a time-periodic electric field.}}}

\vspace {3 cm}
\noindent
\centerline{\small{\bf{Fran\c{c}ois Nicoleau}}}\parn \centerline{\vspace {5 mm}}

\centerline{\small{Laboratoire Jean Leray}}\parn \centerline{\small{UMR CNRS-UN 6629}}\parn
\centerline{\small{D{\'e}partement de Math{\'e}matiques}}\parn
\centerline{\small{2, rue de la Houssini{\`e}re  BP 92208 }}\parn \centerline{\small{F-44322 Nantes cedex 03}}
\vspace {0,1cm}\parn \centerline{\small{e-mail : nicoleau@math.univ-nantes.fr}}\parn

\vspace{3 cm} \parn
\begin{abstract}
We consider the time-dependent Hamiltonian $H(t)=\  {1 \over 2} p^2
-E(t)\cdot x +V(t,x)$ on $L^2(\r^n)$,
where the external electric field $E(t)$ and the short-range electric potential $V(t,x)$
are time-periodic with the same period. It is well-known that the short-range notion depends on the mean value
$E_0$ of the external electric field. When $E_0 = 0$, we show that the high energy limit of the scattering
operators determines uniquely $V(t,x)$. When $E_0 \not= 0$, the same result holds in dimension $n \geq 3$
for generic short-range potentials. In dimension $n= 2$, one has to assume a stronger decay on the electric potential.

\end{abstract}

\vspace{1 cm}
\section{\bf{Introduction.}}

\parn
In this note, we study an inverse scattering problem for a two-body short-range system in the presence of an
external time-periodic electric field $E(t)$ and a time-periodic short-range potential $V(t,x)$ (with the same period $T$).
For the sake of simplicity, we assume that the period $T=1$. \vspace{0,1cm}\parn
The corresponding  Hamiltonian is given on $L^2(\r^n)$ by :
$$ H (t)=\  {1 \over 2} p^2 -E(t) \cdot x +V(t,x) , \num (1.1) $$ 
where $p = -i \partial_x$. When $E(t) =0$, the
Hamiltonian $H(t)$ describes the dynamics of the hydrogen atom placed in a linearly polarized monochromatic
electric field, or a light particle in the restricted three-body problem in which two other heavy particles are
set on prescribed periodic orbits.
When $E(t)=\  \cos (2\pi t) \ E$ with $E \in \r^n,$ the Hamiltonian describes the well-known AC-Stark effect
in the $E$-direction [7].

\vspace{0,5 cm}\parn
In this paper, we assume that the external electric field $E(t)$ satisfies :
$$ t \rightarrow E(t) \ \in L_{loc}^1 (\r; \r^n ) \ \ ,\  \ E(t+1)= E(t) \ a.e \ . \num (A_1) $$
Moreover, we assume that the potential $V \in C^{\infty} (\r \times \r^n )$, is time-periodic
with period $1$, and satisfies the following estimations : $$ \forall \alpha \in \n^n, \ \forall k \in \n,\
\mid \partial_t^k \partial_x^{\alpha} V(t,x) \mid \ \leq \ C_{k,\alpha} \ <x>^{-\delta - \mid \alpha \mid}, \
{\rm{with}} \ \delta > 0, \num (A_2) $$ where $<x> =\  (1 + x^2 )^{1 \over 2}$.
Actually, we can accommodate more
singular potentials
(see [10], [11], [12] for example) and we need $(A_2)$ for only $k, \ \alpha$ with finite order .
It is well-known that under assumptions $(A_1)-(A_2)$, $H(t)$ is essentially self-adjoint on ${\cal S} (\r^n )$
the Schwartz space, [16]. We denote  $H (t)$ the self-adjoint realization with domain $D(H(t))$.

\vspace{0,5 cm}
\parn
Now, let us recall some well-known results in scattering theory for time-periodic electric fields. We denote $H_0 (t)$ the free Hamiltonian :
$$ H_0 (t)= {1 \over 2} p^2 -E(t) \cdot x \ , \num (1.2) $$ and let $U_0(t,s)$, (resp. $U(t,s)$) be the
unitary propagator associated with $H_0 (t)$, (resp. $H(t)$) (see section 2 for details).
\vspace{0,2 cm}\parn For short-range potentials, the wave operators are defined for $s\in \r$ and
$\Phi \in L^2(\r^n)$ by : $$ W^{\pm} (s) \  \Phi  =\  \lim_{t \rightarrow \pm \infty}\ U(s,t)\ U_0 (t,s)\ \Phi .
\num (1.3) $$ We emphasize that the short-range condition depends on the value of the mean of the external
electric field : $$ E_0 \ =\  \int_0^1 \ E(t) \ dt \ . \num (1.4) $$

\vspace{0,2 cm}\parn
$\bullet$
{\bf{The case $E_0 =0$.}} \vspace{0,2 cm}
\parn
By virtue of the Avron-Herbst formula (see section 2), this case falls under the category
of two-body systems with time-periodic potentials and this case was studied by Kitada and Yajima ([10], [11]),
Yokoyama [22].

\vspace{0,2cm}\parn
We recall that for a unitary or self-adjoint operator $U$,
${\cal H}_{c} (U), \ {\cal H}_{ac} (U), \ {\cal H}_{sc} (U)$ and ${\cal H}_{p} (U)$ are, respectively,
continuous, absolutely continuous, singular continuous and point spectral subspace of $U$.

\vspace{0,2cm}\parn
We have the following result ([10], [11], [21]) :

\begin{th}
\hfil\break
Assume that hypotheses $(A_1),\ (A_2)$ are satisfied  with $\delta>1$ and with $E_0 =\ 0$.
\parn \vspace{0,1 cm}
Then : (i) the wave operators $W^{\pm} (s)$ exist for all $s \in \r$.
\parn \vspace{0,1 cm}\hspace{1,2 cm}
(ii) $W^{\pm} (s+1) = W^{\pm} (s)$ and $U(s+1,s)\  W^{\pm} (s)= W^{\pm} (s)\  U_0 (s+1,s)$.
\parn \vspace{0,1 cm}\hspace{1,2 cm}
(iii) $Ran \ (W^{\pm} (s)) =  {\cal H}_{ac} \left( U(s+1, s)\right)$
and
${\cal H}_{sc} \left( U(s+1,s)\right) =  \emptyset$.
\parn \vspace{0,1 cm}\hspace{1,2 cm}
(iv) the purely point spectrum $\sigma_p (U(s+1, s))$ is discrete outside $\{1\}$. \end{th}

\vspace{0,2cm}\parn
{\bf{Comments.}}
\vspace{0,1 cm}\parn
1 - The unitary operators $U(s+1, s)$ are called the
Floquet operators and they are mutually equivalent.  The Floquet operators play a central role in the
analysis of time periodic systems. \parn The eigenvalues of these operators are called Floquet multipliers.
In [5], Galtbayar, Jensen and Yajima improve assertion $(iv)$ :
for $n=\ 3$ and $\delta >2$, ${\cal H}_{p}
\left( U(s+1, s)\right)$ is finite dimensional.

\vspace{0,3 cm}
\parn
 2 - For general $\delta>0$, $W^{\pm}(s)$ do not exist and we have to define other wave operators.
 In ([10], [11]),  Kitada and Yajima have constructed modified wave operators $W_{HJ}^{\pm}$ by solving an
 Hamilton-Jacobi equation.

\vspace{0,8 cm}\parn
$\bullet$
{\bf{The case $E_0 \not=\ 0$.}} \vspace{0,2 cm}
\parn
This case was studied by Moller [12] :  using the Avron-Herbst formula, it suffices to examine
Hamiltonians with a constant external electric field, (Stark Hamiltonians) : the spectral and the scattering
theory for Stark Hamiltonians are well established [2].
In particular, a Stark Hamiltonian with a potential $V$
satisfying $(A_2)$ has no eigenvalues [2].
The following theorem, due to Moller,  is a time-periodic version of these results.

\vspace{0,2 cm}\parn

\begin{th}
\hfil\break
Assume that hypotheses $(A_1), \ (A_2)$ are satisfied with $\delta>{1 \over 2}$ and with $E_0 \not=\ 0$.
\parn \vspace{0,1 cm}
Then : (i) the Floquet operators $U(s+1,s)$ have purely absolutely continuous spectrum.
\parn \vspace{0,1 cm}\hspace{1,2 cm}
(ii) the wave operators $W^{\pm} (s)$ exist for all $s \in \r$ and are unitary.
\parn \vspace{0,1 cm}\hspace{1,2 cm}
(iii) $W^{\pm} (s+1) =W^{\pm} (s)$ and $U(s+1,s)\  W^{\pm} (s)= W^{\pm} (s) \ U_0 (s+1,s)$.
\end{th}

\vspace{1cm}\parn
{\bf{The inverse scattering problem.}}
\vspace{0,2 cm}\parn
For $s \in \r$, we define the scattering operators $S(s)=  W^{+*} (s) \
W^- (s)$.
It is clear that the scattering operators $S(s)$ are periodic with period $1$.

\vspace{0,2 cm}\parn
The inverse scattering problem consists to reconstruct the perturbation $V(s,x)$ from the scattering
operators $S(s)$, $s \in [0,1]$.

\vspace{0,3cm}\parn
In this paper, we prove the following result :

\newpage

\begin{th}
\hfil\break
Assume that $E(t)$ satisfies $(A_1)$ and let $V_j , \ j=\ 1,2$ be potentials satisfying $(A_2)$.
We assume that $\delta>1$ (if $E_0 = 0$), $\delta > {1 \over 2}$ (if $E_0 \not= 0$ and
$n \geq 3$), $\delta >{3 \over 4}$ (if $E_0 \not= 0$
and
$n=\ 2$).
Let $S_j (s)$ be the corresponding scattering operators.
\vspace{0,1 cm}
\parn
Then :
$$
\forall s \in [0,1], \ S_1 (s) =   S_2 (s) \ \Longleftrightarrow \ V_1
=\
V_2 \ .
$$
\end{th}

\vspace{0,5cm}\parn
We prove Theorem 3 by studying  the high energy limit of $[S(s),p]$, (Enss-Weder's approach [4]). We need $n \geq 3$
in the case $E_0 \not= 0$ in order to use the inversion of the Radon transform [6]
on the orthogonal hyperplane to $E_0$.
See also [15] for a similar problem with a Stark Hamiltonian.

\vspace{0,2cm}\parn
We can also remark that if we know the free propagator $U_0 (t,s)$ , $s,t \in \r$,  then by virtue of the following relation :
$$
S(t) = U_0 (t,s) \ S(s) \ U_0 (s,t) \ ,
\num (1.5)
$$
the potential $V(t,x)$ is uniquely reconstructed from the scattering operator $S(s)$ at only
one initial time.

\vspace{0,2cm}\parn
In [21], Yajima proves uniqueness for the case of time-periodic potential with the condition
$\delta > {n \over 2}+1$ and with $E(t)= 0$ by studying the scattering matrices in a high energy regime.
\vspace{0,1cm}\parn
In [20], for a time-periodic potential that decays exponentially at infinity, Weder proves uniqueness 
at a fixed quasi-energy. 
\vspace{0,1cm}\parn
Note also that inverse scattering for long-range time-dependent potentials without external electric fields
was studied by Weder [18] with the Enss-Weder time-dependent method, 
and by Ito for time-dependent electromagnetic
potentials for Dirac equations [8].

\section{\bf{Proof of Theorem 3.}}
\subsection{The Avron-Herbst formula.}
First, let us recall some basic definitions for time-dependent Hamiltonians. Let $\{H(t)\}_{t \in \r}$
be a family of selfadjoint operators on  $L^2(\r^n)$ such that ${\cal{S}}(\r^n ) \subset D(H(t))$
for all $t \in \r$.

\vspace{0,5 cm}\parn
{\bf{Definition.}}
\vspace{0,1 cm}\parn
We call {\it{propagator}} a family of unitary operators on $L^2 (\r^n )$, $U(t,s), \ t,s \in \r$ such that :
\vspace{0,3 cm} \par 1 - $U(t,s)$ is a strongly continuous fonction of $(t,s) \in \r^2$. \par 2 - $U(t,s) \ U(s,r) =
\  U(t,r)$ for all $t,s,r \in \r$. \par 3 - ${\displaystyle{U(t,s) \left( {\cal{S}}(\r^n )\right) \subset
{\cal{S}}(\r^n) }}$ for all $t,s \in \r$.
\par
4 - If $\Phi \in {\cal{S}}(\r^n )$, $U(t,s) \Phi$ is continuously differentiable  in $t$ and $s$ and satisfies :
$$ i\ { {\partial} \over {\partial t}}  U(t,s) \ \Phi =\  H(t)\  U(t,s) \ \Phi \ , \ \ i\
{ {\partial} \over {\partial s}} U(t,s) \ \Phi =\  -U(t,s) \ H(s)
\
\Phi \ .
$$

\vspace{0,5 cm}\parn
To prove the existence and the uniqueness of the propagator for our Hamiltonians $H(t)$, we use a generalization of the
Avron-Herbst formula close to the one given in [3].
\parn
In [12], the author gives, for $E_0 \not= 0$, a different formula which has the
advantage to be time-periodic. To study our inverse scattering problem, we use here a different one, which
is defined for all $E_0$. We emphasize that with our choice, $c(t)$ (see below for the definition of $c(t)$)
is also periodic with period $1$; in particular $c(t)= O(1)$.

\vspace{0,5 cm}\parn
The basic idea is to generalize the well-known Avron-Herbst formula for a Stark Hamiltonian
with a constant electric field $E_0$, [2]; if we consider the Hamiltonian $B_0$ on $L^2 (\r^n)$,
$$
B_0 =\  {1 \over 2} p^2 - E_0 \cdot x \ ,
\num (2.1)
$$
we have the following formula :
$$
e^{-it B_0} = e^{-i { {E_0^2} \over 6} t^3} \ e^{it E_0 \cdot x} \
e^{-i { {t^2} \over 2} E_0 \cdot p} \ e^{-it { {p^2} \over 2}} \ .
\num (2.2)
$$

\vspace{0,5 cm}\parn
In the next definition, we give a similar formula for time-dependent electric fields.

\vspace{0,5 cm}\parn
{\bf{Definition.}}
\vspace{0,1 cm}\parn
We consider the family of unitary operators $T(t)$, for $t \in \r$ : $$
T(t) =\  e^{-ia(t)} \ e^{-ib(t)\cdot x} \ e^{-ic(t)\cdot p} \ , $$ where : $$
b(t) =\  - \int_0^t \ (E(s) -E_0) \ ds - \int_0^1 \int_0^t \ (E(s) -E_0)
\
ds \ dt \ .
\num (2.3)
$$
$$
c(t) =\  -\int_0^t \ b(s) \ ds \ .
\num (2.4)
$$
$$
a(t) =\  \int_0^t \ \left( {1 \over 2} \ b^2 (s) -E_0 \cdot c(s) \right)
\
ds \ .
\num (2.5)
$$

\vspace{0,5 cm}

\begin{lem}
\hfil\break
The family $\{H_0 (t)\}_{t \in \r}$ has an unique propagator $U_0 (t,s)$ defined by :
$$ U_0 (t,s) =\ T(t) \ e^{-i(t-s) B_0} \ T^* (s) \ .
\num (2.6)
$$
\end{lem}

\vspace{0,5 cm}\parn
{\bf{Proof.}}
\vspace{0,1 cm}\parn
We can always assume $s=\ 0$ and we make the following ansatz :
$$
U_0 (t,0) =\  e^{-ia(t)} \ e^{-ib(t)\cdot x}
\ e^{-ic(t)\cdot p} \ e^{-it B_0} \ .
\num (2.7)
$$
Since on the Schwartz space, $U_0 (t,0)$ must satisfy :
$$
i\ { {\partial} \over {\partial t}} \ U_0(t,0)  =  H_0 (t)\  U_0(t,0) \ ,
\num (2.8)
$$
the functions $a(t), \ b(t), c(t)$ solve :
$$
\dot{b}(t) =  -E(t)+E_0, \ \dot{c}(t) =  -b(t), \ \dot{a}(t) =  {1 \over 2} \ b^2 (t) - E_0 \cdot c(t) .
\num (2.9)
$$
We refer to [3] for details and [12] for a different formula. $\Box$

\vspace{0,8 cm}\parn
In the same way, in order to define the propagator corresponding to the family $\{H(t)\}$,
we consider a Stark Hamiltonian with a time-periodic potential $V_1 (t,x)$, (we recall that $c(t)$ a is
$C^1$-periodic function) :
$$
B(t) =\  B_0 + V_1(t, x) \ \ {\rm{where}} \ \ V_1 (t,x)=\ e^{i c(t)
\cdot p} \ V(t,x) \ e^{-i c(t) \cdot p} =\  V(t,x+c(t)) .
\num (2.10)
$$
Then, $B(t)$ has an unique propagator $R(t,s)$, (see [16] for the case $E_0=0$ and [12] for
the case $E_0 \not=0$). It is easy to see that the propagator $U(t,s)$ for the family $\{H(t)\}$ is defined by :
$$
U(t,s) =\  T(t) \ R(t,s) \ T^* (s) .
\num (2.11)
$$

\vspace{0,5 cm}
\parn
{\bf{Comments.}}
\vspace{0,2 cm}\parn
Since the Hamiltonians $H_0(t)$ and $H(t)$ are time-periodic with period $1$, one has for all $t,s \in \r$  :
$$
U_0(t+1,s+1) = U_0(t,s) \ \ ,\ \ U(t+1,s+1) = U(t,s) \ .
\num (2.12)
$$
Thus, the wave operators satisfy $W^{\pm} (s+1) = W^{\pm} (s)$.

\vspace{0,5 cm}\parn
\subsection{The high energy limit of the scattering operators.}

\vspace{0,4 cm}
\parn
In this section, we study the high energy limit of the scattering operators by using the well-known Enss-Weder's
time-dependent method [4]. This method can be used to study Hamiltonians with electric and magnetic potentials on
$L^2 (\r^n )$ [1], the Dirac equation [9], the N-body case [4], the Stark effect ([15], [17]),
the Aharonov-Bohm effect [18].

\vspace{0,3 cm}\parn In [13], [14]  a stationary approach, based on the same ideas, is proposed to solve scattering inverse problems for
Schr\"odinger operators with magnetic fields or with the Aharonov-Bohm effect.

\newpage
\parn
Before giving the main result of this section, we need some notation. \vspace{0,5 cm} \par
$\bullet$ $\Phi, \Psi$
are the Fourier transforms of functions in $C_0^{\infty} (\r^n )$. \par
$\bullet$ $ \omega \in S^{n-1} \cap \Pi_{E_0}$
is fixed, where $\Pi_{E_0}$ is the orthogonal hyperplane to $E_0$. \par
$\bullet$ $\fta =\  \ex \Phi, \ \ftb =\  \ex \Psi$. \par

\vspace{0,5 cm}
\parn
We have the following high energy asymptotics  where $<\ , \ >$ is the usual scalar product in $L^2 (\r^n )$ :

\vspace{0,1cm}\parn

\begin{prop}
\hfil\break Under the assumptions of Theorem 3, we have for all $s \in [0,1] $,
$$ < [S(s), p] \ \fta \ , \ \ftb  > \ =\
{\lambda}^{-{1 \over 2}} \ < \left( \int_{-\infty}^{+\infty} \
\partial_x V(s,x+t\omega) \ dt \right) \ \Phi \ ,\ \Psi > + o\ (
{\lambda}^{-{1 \over 2}}) \ .
$$
\end{prop}

\vspace{0,5 cm}
\parn
{\bf{Comments.}}
\vspace{0,1 cm}\parn
Actually, for the case $n=2, \ E_0\not=0$ and $\delta > {3 \over 4}$, Proposition 5 is also valid for
$\omega \in S^{n-1}$ satisfying $\mid \omega \cdot E_0  \mid < \mid E_0 \mid$, (see ([18], [15]).
\vspace{0,1 cm}\parn
Then, Theorem 3 follows from Proposition 5  and the inversion of Radon transform ([6] and [15], Section 2.3).

\vspace{0,5 cm}
\parn
{\bf{Proof of Proposition 5.}}
\vspace{0,2 cm}\parn
For example, let us show Proposition 5 for the case $E_0\not=0$ and $n \geq 3$, the other cases are similar.
More precisely, see [18] for the case $E_0 =0$,  and for the case $n=2, \
E_0 \not=0$, see ([17], Theorem 2.4) and ([15], Theorem 4).

\vspace{0,4 cm}
\parn
{\bf{Step 1.}}
\vspace{0,2 cm}\parn
Since $c(t)$ is periodic,
$c(t)=\ O(1)$. Then,  $V_1(t,x)$ is a
short-range perturbation of $B_0$, and we can define the usual wave operators for the pair of Hamiltonians
$(B(t), B_0)$ :
$$
\Omega^{\pm} (s) =\  {\rm{s}}-\lim_{t \rightarrow \pm \infty} \ R(s,t) \
e^{-i(t-s) B_0} \ .
\num (2.13)
$$
Consider also the scattering operators
$S_1 (s) =\  \Omega^{+*} (s)\ \Omega^{-} (s)$. By virtue of $(2.6)$ and
$(2.11)$, it is clear that :
$$
S(s) =  T(s) \ S_1 (s) \ T^*(s) \ .
\num (2.14)
$$
Using the fact that $e^{-ib(s)\cdot x} \ p\ e^{ib(s)\cdot x} = p+b(s)$, we have :
$$
[S(s), p] = [S(s), p+b(s)] =  T(s) \ [S_1(s), p]\ T^* (s) \ .
\num (2.15)
$$
Thus,
$$ < [S(s), p] \ \fta \ , \ \ftb  > = \  < [S_1(s), p] \ T^* (s) \ \fta \ , \ T^* (s)\ \ftb  >.
\num (2.16)
$$
In other hand,
$$
T^* (s) \ \fta =\  \ex \  e^{i c(s)\cdot (p+{\sqrt{\lambda}} \omega)} \
e^{ib(s) \cdot x} \ e^{ia(s)} \ \Phi.
\num (2.17)
$$
So, we obtain :
$$
< [S(s), p] \ \fta \ , \ \ftb  > \ =\  < [S_1(s), p] \ \Fta, \
\Ftb \ >,
\num (2.18)
$$
where
$$
f =\    e^{ic(s) \cdot p} \ e^{ib(s) \cdot x} \ \Phi  \ {\rm{and}} \  g =\    e^{ic(s) \cdot p}
\ e^{ib(s) \cdot x} \ \Psi \ .
\num (2.19)
$$
Clearly, $f, \ g$ are the Fourier transforms of functions in $C_0^{\infty} (\r^n )$.

\vspace{0,5 cm} \parn
$\bullet$ { \bf{Step 2 : Modified wave operators.}}
\vspace{0,2 cm}\parn
Now, we follow a strategy close to [15] for time-dependent potentials.
First, let us define a free-modified dynamic $U_D (t,s)$ by :
$$
U_D(t,s) = e^{-i(t-s) B_0} \ e^{-i \int_0^{t-s} \ V_1 (u+s, up' + {1 \over 2} u^2 E_0 ) \ du} \ ,
\num (2.20)
$$
where $p'$ is the projection of $p$ on the orthogonal hyperplane to $E_0$.
\vspace{0,2 cm}\parn
We define the modified wave operators :
$$
\Omega_D^{\pm} (s) = {\rm{s}}-\lim_{t \rightarrow \pm \infty} \ R(s,t)
\ U_D (t,s) \ .
\num (2.21)
$$
It is clear that :
$$
\Omega_D^{\pm}(s) = \Omega^{\pm}(s) \ e^{-i g^{\pm} (s,p')} \ ,
\num (2.22)
$$
where
$$
g^{\pm} (s,p') \ =\ \int_0^{\pm \infty} \ V_1 (u+s, up'+ {1 \over 2} u^2 E_0 ) \ du \ .
\num (2.23)
$$
Thus, if we set
$S_D (s)= \Omega_D^{+*}(s) \Omega_D^- (s)$, one has  :
$$
S_1 (s) = e^{-i g^{+} (s,p')} \ S_D (s) \ e^{i g^{-} (s,p')}
\num (2.24)
$$

\vspace{0,5 cm} \parn
$\bullet$ { \bf{Step 3 : High energy estimates.}}
\vspace{0,3 cm}\parn
Denote $\rho = min \ (1, \delta)$. We have the following estimations,
(the proof is exactly the same as in ([15], Lemma 3)  for time-independent potentials).

\vspace{0,2 cm}\parn
\begin{lem}
\hfil\break
For $\lambda >>1$, we have :
$$
\mid \mid \ \left( V_1(t,x) - V_1(t, (t-s)p'+ {1 \over 2} (t-s)^2 E_0) \right)  \ U_D (t,s) \  e^{i g^{\pm} (s,p') }
\Fta \ \mid \mid \ \hspace{3 cm}
\num (i)
$$
$$
\hspace{10 cm} \leq \ C \ (1 +\mid (t-s) {\sqrt \lambda} \mid )^{- {1 \over 2} - \rho }  \ .
$$
$$
\mid \mid \ \left( R(t,s) \Omega_D^{\pm}(s) -U_D (t,s)\right) e^{i g^{\pm} (s,p')} \Fta \ \mid \mid \ =
\ O \ ( {\lambda}^{-{1 \over 2}} ) \ , \ {\rm{uniformly \  for}}\  t, \ s \in \r \ .
\num (ii)
$$
\end{lem}

\vspace{0,5 cm} \parn
$\bullet$ { \bf{Step 4.}}
\vspace{0,2 cm}\parn

\vspace{0,1 cm} \parn
We denote $F(s, \lambda, \omega) = < [S_1 (s), p] \ \Fta \ , \ \Ftb  >$.
Using $(2.24)$, we have :
$$
F(s,\lambda, \omega) \ = \ < [ e^{-i g^{+} (s,p')} \ S_D (s) \ e^{i g^{-} (s,p')} , p] \ \Fta \ , \ \Ftb >
$$
$$
\hspace{ 1,4cm} =  \ <[S_D (s),p] \   e^{i g^{-} (s,p')}\Fta \ , \ e^{i g^{+} (s,p')} \Ftb >
$$
$$
 \hspace{ 3,5cm}= \ <[S_D (s)-1,p-\la \omega ]\   e^{i g^{-} (s,p')}\Fta \ , \ e^{i g^{+} (s,p')} \Ftb >
$$
$$
\hspace{ 2,4cm}= \ < (S_D(s)-1) \ e^{i g^{-} (s,p')} \Pta \ , \ e^{i g^{+} (s,p')} \Ftb >
$$
$$
\hspace{ 2,4cm}- \ < (S_D (s)-1) \ e^{i g^{-} (s,p')}\Fta \ , \ e^{i g^{+} (s,p')} \Ptb >
$$
\hspace {4,9 cm}
$:= F_1 (s,\lambda, \omega) - F_2 (s,\lambda, \omega)$.

\vspace{ 1 cm}
\parn
First, let us study $ F_1 (s, \lambda, \omega) $.
 Writing $S_D (s)-1 = (\Omega_D^+ (s) - \Omega_D^- (s))^* \ \Omega_D^- (s)$ and using
 $$ \Omega_D^+ (s) - \Omega_D^- (s)\ = \ i \ \int_{- \infty}^{+ \infty} \ R(s,t) \
\left( V_1 (t,x) - V_1 (t,(t-s)p'+ {1 \over 2} (t-s)^2 E_0 )\right) \ U_D(t,s) \ dt \ ,
\num (2.25)
$$
we obtain :
$$
S_D (s)-1 = - i  \ \int_{- \infty}^{+ \infty} \  U_D(t,s)^*  \
\left( V_1 (t,x) - V_1 (t,(t-s)p'+ {1 \over 2} (t-s)^2 E_0 )\right)
\num (2.26)
$$
$$
\hspace{13 cm}  R(t,s) \ \Omega_D^- (s) \ dt \ .
$$
Thus,
$$ F_1 (s, \lambda, \omega) = - i  \ \int_{- \infty}^{+ \infty} \ < R(t,s) \ \Omega_D^- (s)\
e^{i g^{-} (s,p')}\Pta \ ,
$$
$$
\hspace{5,7 cm} \left( V_1(t,x) - V_1 (t, (t-s)p'+ {1 \over 2} (t-s)^2 E_0 )\right) \ U_D(t,s) \
e^{i g^{+} (s,p')}
\Ftb > \ dt
$$
$$
\hspace {1,2 cm} = - i  \ \int_{- \infty}^{+ \infty} \ <   U_D(t,s)   \
e^{i g^{-} (s,p')} \Pta \ , \
$$
$$
\hspace{5,6 cm} \left( V_1 (t,x) - V_1 (t,(t-s)p'+ {1 \over 2} (t-s)^2 E_0 )\right)
\ U_D(t,s) \ e^{i g^{+} (s,p')} \Ftb > \ dt
$$
\hspace {5,6 cm} $+ \ R_1 (s,\lambda, \omega) \ ,$
\parn
where :
$$
R_1 (s,\lambda, \omega) \ = \ - i  \ \int_{- \infty}^{+ \infty} \ < \left( R(t,s) \ \Omega_D^- (s) - U_D(t,s)
\right) \ e^{i g^{-} (s,p')}\Pta \ ,
\num (2.27)
$$
$$
\hspace {5,5 cm}
\ \left( V_1 (t,x) - V_1 (t,(t-s)p'+ {1 \over 2} (t-s)^2 E_0 )\right)
\ U_D(t,s) \ e^{i g^{+} (s,p')} \Ftb > \ dt \ .
$$

\vspace{0,5 cm}
\parn
By Lemma 6, it is clear that $R_1 (s,\lambda, \omega) = O \  ( \lambda^{-1})$.
Thus, writing $t = { {\tau} \over {\la}}+s $, we obtain :

\newpage
$$
F_1 (s,\lambda, \omega)\ = - { i \over \la}  \ \int_{- \infty}^{+ \infty} \
<  U_D ( \T +s,s) \ e^{i g^{-} (s,p')} \ \Pta \ , \hspace{3 cm}
\num (2.28)
$$
$$
\hspace{4 cm}\left(  V_1(\T+s,x)   - V_1( \T+s, \T p'+ { {\tau^2} \over {2\lambda}}  E_0 )\right) \
$$
$$
\hspace{4,2  cm}U_D (\T +s,s ) \ e^{i g^{+} (s,p')} \Ftb > \ d\tau +\ O \ ( \lambda^{-1}) \ .
$$
Denote by $f_1 (\tau, s, \lambda, \omega )$ the integrand of the (R.H.S) of $(2.28)$. By Lemma 6 (i),
$$
\mid f_1 (\tau, s , \lambda, \omega ) \mid \ \leq C \ (1 + \mid \tau \mid )^{- {1 \over 2} - \rho }  \ .
\num (2.29)
$$
So, by Lebesgue's theorem, to obtain the asymptotics of $F_1 (s,\lambda, \omega)$, it suffices to determine
${\displaystyle{\lim_{\lambda \rightarrow + \infty} \ f_1 (\tau, s , \lambda, \omega )}}$.

\vspace{0,6 cm}
\parn
Let us denote :
$$
U^{\pm} (t,s,p') = e^{i \ \int_t^{\pm \infty} \ V_1(u+s,up' + {1 \over 2} u^2 E_0) \ du}.
\num (2.30)
$$
We have :
$$
f_1 (\tau,s, \lambda, \omega ) \ = \ < e^{-i {\tau \over \la} B_0 } \ U^- ( \T , s, p') \ \Pta \ ,
\hspace{4 cm}
\num (2.31)
$$
$$
\hspace{4 cm} \left( V_1(\T +s,x) - V_1( \T+s, \T p'+  { {\tau^2} \over {2\lambda}} \ E_0 )\right) \
e^{-i {\tau \over \la} B_0 } \ U^+ ( \T,s, p') \ \Ftb > \ .
$$

\vspace{0,3 cm}\parn
Using the Avron-Herbst formula $(2.2)$, we deduce that :
$$
f_1 (\tau, s, \lambda, \omega ) \ = \ < e^{-i {\tau \over {2\la}} p^2 } \ U^- ( \T ,s, p') \ \Pta \ ,
\num (2.32)
$$
$$
\left( V_1(\T+s, x + { {\tau^2} \over {2\lambda}} \ E_0 ) - V_1( \T+s, \T p'+  { {\tau^2} \over {2\lambda}} \
E_0 )\right) \
e^{-i {\tau \over {2\la}} p^2 } \ U^+ ( \T ,s, p') \ \Ftb > \ .
$$

\vspace{0,3 cm}
\parn
Then, we obtain :
$$
f_1 (\tau, s, \lambda, \omega ) \ = \ < e^{-i {\tau \over {2\la}} (p+\la \omega)^2 } \ U^- ( \T ,s,
p'+ \la \omega) \ p f \ , \
\num (2.33)
$$
$$
\hspace{5 cm}\left( V_1(\T+s, x + { {\tau^2} \over {2\lambda}}\ E_0 ) - V_1( \T+s, \T (p'+ \la \omega)
+ { {\tau^2} \over {2\lambda}}\ E_0 )\right) \
$$
$$
\hspace{3 cm}e^{-i {\tau \over {2\la}} (p+\la \omega)^2  } \
U^+ ( \T , s, p'+\la \omega ) \ g > \ .
$$

\vspace{0,5 cm}
\parn
Since
$$
e^{-i {\tau \over {2\la}} (p+\la \omega)^2 } \ = \ e^{-i { {\tau \la}  \over 2}} \ e^{-i\tau \omega.p} \
e^{-i {\tau \over {2\la}} p^2} \ ,
\num (2.34)
$$
we have
$$
f_1 (\tau, s, \lambda, \omega ) \ = \ < e^{-i {\tau \over {2\la}} p^2} \
U^{-} ( \T , s, p'+\la \omega ) \ \ p f \ ,\
\hspace{5,5 cm}
\num (2.35)
$$
$$ \left( V_1(\T+s, x + \tau \omega + { {\tau^2} \over {2\lambda}}\  E_0 ) - V_1(\T+s, \T (p'+\la \omega)
+ { {\tau^2} \over {2\lambda}}\  E_0 )\right)
$$
$$
\hspace{8 cm} e^{-i {\tau \over {2\la}} p^2} \
U^{+} ( \T ,s, p'+\la \omega )\ g > \ .
$$

\vspace{0,2 cm}
\parn
Since
$\mid V_1 (u+s,  u (p'+ \la \omega ) + {1 \over 2 } u^2 E_0 )) \mid \ \leq \ C \ (u^2 +1)^{-\delta} \in L^1 (\r^+ , \ du )$, it is easy to show (using Lebesgue's theorem again) that :
$$
s-\lim_{\lambda \rightarrow + \infty} \ U^{\pm} ( \T ,s, p'+\la \omega ) \ = 1 \ .
\num (2.38) $$

\parn
Then,
$$
\lim_{\lambda \rightarrow + \infty} \ f_1 (\tau, s , \lambda, \omega ) \ = \ < pf \ ,\
\left( V_1 (s, x+ \tau \omega)- V_1 (s, \tau \omega)\right) \ g > \ .
\num (2.39)
$$
So, we have obtained :
$$
F_1 (s, \lambda, \omega) \ =\ -{ {i} \over {\la}} \ <pf, \ \left( \int_{-\infty}^{+\infty} \
(V_1 (s, x+ \tau \omega)- V_1 (s, \tau \omega)) \ d\tau  \right)\  g > \ + o( {1 \over \la}).
\num (2.40)
$$


\vspace{0,5 cm}
\parn
In the same way, we obtain
$$
F_2 (s, \lambda, \omega) \ =\ -{ {i} \over {\la}} \ <f, \ \left( \int_{-\infty}^{+\infty} \
(V_1 (s, x+ \tau \omega)- V_1 (s, \tau \omega)) \ d\tau  \right)\  pg > \ + o( {1 \over \la}) \ ,
\num (2.41)
$$
so
$$
F (s, \lambda, \omega) \ =\ F_1 (s, \lambda, \omega)- F_2(s, \lambda, \omega)
\hspace{0,8 cm}
\num (2.42)
$$
$$
\hspace{4 cm}  =\ { 1 \over {\la}} \ <f, \ \left( \int_{-\infty}^{+\infty} \
\partial_x  V_1 (s, x+ \tau \omega) \ d\tau  \right)\  g > \ + o( {1 \over \la}) \ .
\num (2.43)
$$
Using $(2.19)$ and $\partial_x V(s,x + \tau \omega) = e ^{-i c(s) \cdot p} \
\partial_x  V_1 (s, x+ \tau \omega)\ e ^{i c(s) \cdot p}$, we obtain :
$$
F (s, \lambda, \omega)  \ =\ \ { 1 \over {\la}} \ <\Phi, \ \left( \int_{-\infty}^{+\infty} \
\partial_x  V (s, x+ \tau \omega) \ d\tau  \right)\  \Psi > \ + o( {1 \over \la}) \ . \ \Box
\num (2.44)
$$

\vspace{5mm}
\parn

\end{document}